\documentclass[10pt]{revtex4}
\usepackage{amssymb}
\usepackage{latexsym}                          
\usepackage{epsfig}
\usepackage{amsmath}
\begin{document}

\title{Einstein Equations and MOND Theory from Debye Entropic Gravity}
\author{A.  Sheykhi$^{1,2}$ \footnote{
sheykhi@uk.ac.ir} and K. Rezazadeh Sarab$^{3}$}
\address{$^1$ Center for Excellence in Astronomy and Astrophysics (CEAA-RIAAM) Maragha, P. O. Box 55134-441, Iran\\
         $^2$ Physics Department and Biruni Observatory, Shiraz University, Shiraz 71454, Iran\\
         $^3$ Department of Physics, Shahid Bahonar University, P.O. Box 76175, Kerman, Iran}

\begin{abstract}
Verlinde's proposal on the entropic origin of gravity is based
strongly on the assumption that the equipartition law of energy
holds on the holographic screen induced by the mass distribution
of the system. However, from the theory of statistical mechanics
we know that the equipartition law of energy does not hold in the
limit of very low temperature. Inspired by the Debye model for the
equipartition law of energy in statistical thermodynamics and
adopting the viewpoint that gravitational systems can be regarded
as a thermodynamical system, we modify Einstein field equations.
We also perform the study for Poisson equation and modified
Newtonian dynamics (MOND). Interestingly enough, we find that the
origin of the MOND theory can be understood from Debye entropic
gravity perspective. Thus our study may fill in the gap existing
in the literature understanding the theoretical origin of MOND
theory. In the limit of high temperature our results reduce to
their respective standard gravitational equations.

\textit{keywords:} entropic; gravity; Debye model.

\end{abstract}
 \maketitle

\bigskip
\section{Introduction}
Thermodynamics of black holes reveals that geometrical quantities
such as horizon area and surface gravity are related to the
thermodynamic quantities such as entropy and temperature. The
first law of black hole thermodynamics implies that the entropy
and the temperature together with the energy (mass) of the black
hole satisfy $dE=TdS$ \cite{HB}. In $1995$ Jacobson \cite {Jac}
put forwarded a new step and suggested that the hyperbolic second
order partial differential Einstein equation for the spacetime
metric has a predisposition to thermodynamic behavior. He
disclosed that the Einstein field equation is just an equation of
state for the spacetime and in particular it can be derived from
the proportionality of entropy and the horizon area together with
the fundamental relation $\delta Q=TdS$. Following Jacobson,
however, several recent investigations have shown that there is
indeed a deeper connection between gravitational dynamics and
horizon thermodynamics. The deep connection between horizon
thermodynamics and gravitational dynamics, help to understand why
the field equations should encode information about horizon
thermodynamics. These results prompt people to take a statistical
physics point of view on gravity.

A next great step put forwarded by Verlinde \cite{Verlinde} who claimed
that the laws of gravity are not fundamental and in particular
they emerge as an entropic force caused by the changes in the
information associated with the positions of material bodies.
According to Verlinde proposal when a test particle with mass $m$
approaches a holographic screen from a distance $\triangle x$, the
change of entropy on the holographic screen is
\begin{equation}\label{deltaS}
\triangle S=2\pi \frac{m}{\hbar} \triangle x,
\end{equation}
where we have set  $k_{B}=c=1$ for simplicity, through this paper.
The entropic force can arise in the direction of increasing
entropy and is proportional to the temperature,
\begin{equation}\label{entropic force}
F=T \frac{\triangle S}{\triangle x}.
\end{equation}
Verlinde's derivation of Newton's law of gravitation at the very
least offers a strong analogy with a well understood statistical
mechanism. Therefore, this derivation opens a new window to
understand gravity from the first principles. The study on the
entropic  force has raised a lot of attention recently (see
\cite{Cai4,Other,newref,sheyECFE,Ling,Modesto,Yi,Sheykhi2} and
references therein).

Verlinde's proposal on the entropic origin of gravity is based
strongly on the assumption that the equipartition law of energy
holds on the holographic screen induced by the mass distribution
of the system, namely, $E=\frac{1}{2} N T$. However, from the
theory of statistical mechanics we know that the equipartition law
of energy does not hold in the limit of very low temperature. By
low temperature, we mean that the temperature of the system is
much smaller than Debye temperature, i.e. $T\ll T_D$. It was
demonstrated that the Debye model is very successful in
interpreting the physics at the very low temperature. Hence, it is
expected that the equipartition law of energy for the
gravitational systems should be modified in the limit of very low
temperature (or very weak gravitational field).

It is important to note that Verlinde got the Newton's law of
gravitation, Einstein equations and Poisson equation with the
assumption that each bit on holographic screen is free of
interaction. It should be more general  that the bits on
holographic screen interact each others. In such case, one could
anticipate that the Newton's law of gravitation, Einstein equation
and Poisson equation must be modified. For example, Gao \cite{Gao}
studied  three dimensional Debye model and modified the entropic
force and henece Friedmann equations. Such modification can
interpret the current acceleration of the universe without
invoking any kind of dark energy \cite{Gao}. In this paper we use
the Debye model to modify the entropic gravity. We find that this
modified entropic force affects on the law of gravitations and
modify them accordingly.

This paper is structured as follows.  In the next section we
derive Einstein field equations from Debye entropic gravity. The
theoretical origin of MOND theory is discussed in the framework of
Debye entropic gravity in section III. Sec. IV is devoted to the
derivation of the Poisson equation from Debye entropic force
scenario. We finish our paper with conclusions which appear in
Sec. V.

\section{Einstein equations from Debye entropic gravity}
Following Verlinde's scenario, gravity may have a statistical
thermodynamics origin. Thus, any modification of statistical
mechanics should modify the laws of gravity accordingly. In this
section we use the modified equipartition law of energy to obtain
the modified Einstein equations.

We consider a system that its boundary is not infinitely extended
and forms a closed surface. We can take the boundary as a storage
device for information, i.e. a holographic screen. Assuming that
the holographic principle holds, the maximal storage space, or
total number of bits $N$, is proportional to the area $A$,
\begin{equation} \label{N&A}
N= \frac{A}{G \hbar}.
\end{equation}

Suppose there is a total energy $E$ present in the system. Let us
now just make the simple assumption that the energy is divided
evenly over the bits. Each bit on the holographic screen has one
dimensional degree of freedom, hence we can use the one
dimensional equipartition law of energy. The equipartition law of
energy which is valid in all range of temperatures is
\begin{equation}\label{E}
E=\frac{1}{2} N T D(x),
\end{equation}
where T is the temperature of the screen and $D(x)$ is the one
dimensional Debye function defined as
\begin{equation}\label{Dx}
D(x) \equiv \frac{1}{x} \int_0^x \frac{y}{e^y -1 }dy,
\end{equation}
and $x$ is related to the temperature
\begin{equation}\label{x}
x \equiv \frac{T_D}{T},
\end{equation}
where $T_D$ is the Debye temperature. Using the equivalence
between mass and energy, $E=M$, as well as Eq. (\ref{N&A}), we can
rewrite Eq. (\ref{E}) in a more general form,
\begin{equation}\label{M&T}
M=\frac{1}{2G \hbar} \oint_S T D(x) dA,
\end{equation}
where the integration is over the holographic screen. For
temperature, we use the Unruh temperature formula on the
holographic screen,
\begin{equation}\label{TUnruh}
T=\frac{\hbar a}{2 \pi},
\end{equation}
where $a$ denotes the acceleration. The acceleration has relation
with the Newton's potential and in general relativity it may be
written as
\begin{equation} \label{a&phi1}
a^b=-\nabla^b \phi,
\end{equation}
where $\phi$ is the natural generalization of Newton's potential
in general relativity and for it we have \cite{Wald},
\begin{equation} \label{phi&xi}
\phi =\frac{1}{2} Ln(-\xi^a \xi_a),
\end{equation}
where $\xi^a$ is a global time like Killing vector. {The exponent
$e^\phi$ represents the redshift factor that  relates the local
time coordinate to that at a reference point with $\phi=0$, which
we will take to be at infinity. We choose the holographic screen
$S$ as a closed equipotential surface or in other words, a closed
surface of constant redshit $\phi$. Therefore Eq. (\ref{TUnruh})
may be written as \cite {Verlinde}
\begin{equation}\label{T}
T=\frac{\hbar}{2 \pi}e^\phi N^a \nabla_a \phi,
\end{equation}
where $N^a$ is the unit outward pointing vector that is normal to
the equipotential holographic screen $S$ and time like Killing
vector $\xi^b$. We inserted a redshift factor $e^\phi$, because
the temperature $T$ is measured with respect to the reference
point at infinity. Because $N^a$ is normal to the equipotential
holographic screen, for it we have
\begin{equation} \label{Na}
N^a=\frac{\nabla^a \phi}{(\nabla^b \phi \nabla_b \phi)^{1/2}}.
\end{equation}
Therefore we can rewrite Eq. (\ref{T}) as
\begin{equation} \label{T&phi1}
T=\frac{\hbar}{2\pi}e^\phi (\nabla^a\phi \nabla_a\phi)^{1/2}.
\end{equation}
} { Substituting Eq. (\ref{T}) in Eq. (\ref{M&T}), we  get
\begin{equation} \label{M&phi}
M=\frac{1}{4 \pi G} \oint_S e^{\phi} N^a \nabla_a \phi D(x) dA.
\end{equation}
Following the same logic of \cite{Wald}, we can obtain
\begin{equation} \label{M&xi}
M=-\frac{1}{8 \pi G} \oint_S \nabla^a \xi^b D(x) dS_{ab},
\end{equation}
where $dS_{ab}$ is the two-surface element \cite{Poisson}. On the
other hand, according to the Stokes theorem, we have
\cite{Poisson}
\begin{equation} \label{Stokes}
\oint_S B^{ab} dS_{ab}=2 \int_\Sigma \nabla_b B^{ab} d\Sigma_a,
\end{equation}
where $B^{ab}$ is an antisymmetric tensor field and $S$ is the two dimensional boundary
of the hypersurface $\Sigma$. $d\Sigma_a$ is a directed surface element on $\Sigma$ and for it we have
\begin{equation} \label{dSigma&n}
d\Sigma_a=\varepsilon n_a d\Sigma,
\end{equation}
where $n^a$ is the unit normal of the hypersurface $\Sigma$ and
$\varepsilon$ is equal to -1 or 1 if the hypersurface is spacelike
or timelike, respectively.} { Now we apply the Stokes theorem
(\ref{Stokes}) for Eq. (\ref{M&xi}) and get
\begin{align} \label{M&nabla}
\nonumber
M&=-\frac{1}{4 \pi G} \int_\Sigma \nabla_b[\nabla^a \xi^b D(x)] d\Sigma_a
\\ \nonumber
&=-\frac{1}{4 \pi G} \int_\Sigma [ D(x)\nabla_b \nabla^a \xi^b
+\nabla^a \xi^b \nabla_b D(x)]d\Sigma_a
\\
&=-\frac{1}{4 \pi G} \int_\Sigma [-D(x)\nabla_b \nabla^b \xi^a
+\nabla^a\xi^b\nabla_b D(x)] d\Sigma_a,
\end{align}
where in the last step we have used the Killing equation,
\begin{equation} \label{Killing}
\nabla^a \xi^b+\nabla^b \xi^a=0.
\end{equation}
Now we use the relation \cite{Wald}
\begin{equation} \label{xi&Rab}
\nabla^a \nabla_a \xi^b=-R^b_a\xi^a,
\end{equation}
which is implied by the Killing equation for $\xi^a$, and get
\begin{align}\label{M&Rab}
\nonumber
M &= -\frac{1}{4\pi G}\int_\Sigma [R_{ab} \xi^b D(x)+ \nabla_a \xi^c \nabla_c D(x)] d\Sigma^a
\\ \nonumber
&=-\frac{1}{4 \pi G}\int_\Sigma [R_{ab}\xi^b D(x) + e^{-2\phi} (-\xi^b \xi_b)\nabla_a \xi^c \nabla_c D(x)] d\Sigma^a
\\
&=\frac{1}{4 \pi G}\int_\Sigma [R_{ab} D(x) - e^{-2\phi} \xi_b \nabla_a \xi^c \nabla_c D(x)] n^a \xi^b d\Sigma,
\end{align}
where in the second line we have used Eq. (\ref{phi&xi}). In the
last line we have used $d\Sigma^a=-n^a d\Sigma$, because the
hypersurface $\Sigma$ is spacelike. }

{ On the other hand, $M$ can be expressed as an integral over the
enclosed volume of certain components of stress energy tensor
$\mathcal{T}_{ab}$ \cite{Wald},
\begin{equation}\label{M&Tab}
M=2 \int (\mathcal{T}_{ab} -\frac{1}{2} \mathcal {T} g_{ab})n^a
\xi^b d\Sigma.
\end{equation}
Equating Eqs. (\ref{M&Rab}) and (\ref{M&Tab}), we find
\begin{equation}\label{modifiedEinstein}
D(x) R_{ab}- e^{-2\phi}\xi_b \nabla_a \xi^c\nabla_c D(x)=8 \pi G
(\mathcal{T}_{ab}-\frac{1}{2}\mathcal{T} g_{ab}).
\end{equation}
The above equation is the modified Einstein equations resulting
from considering the Debye correction to the equipartition law of
energy in the framework of entropic gravity scenario. This
equation is now valid for all range of temperature, since we have
assumed the general equipartition law of energy.  Therefore, we
see that in Verlinde's approach, any modification of first
principles such as equipartition law of energy will modify the
gravitational field equations. The question whether the modified
term in Einstein equation can be detectable practically or not
needs more investigations in the future. One needs to first
specify the Debye function $ D(x)$ and then try to solve the field
equations (\ref{modifiedEinstein}). The resulting solutions should
be checked with experiments or observations. It is clear that the
correction term only plays role in very low temperature, in which
the curvature of spacetime tends to zero and it becomes flat.}

{It is instructive to examine the modified Einstein equations in
the high temperatures limit. According to the Unruh temperature
formula we have
\begin{equation} \label{g&T}
g=\frac{2\pi}{\hbar}T,
\end{equation}
where $g$ is the norm of the gravitational acceleration.
Therefore, the strength of the gravitational field is proportional
to the temperature. Also, we can define the Debye acceleration
relating to the Debye temperature as
\begin{equation} \label{gD&TD}
g_D=\frac{2\pi}{\hbar}T_D.
\end{equation}
Therefore, if the temperature is larger than the Debye
temperature, i.e. $T>T_D$, then the norm of the gravitational
acceleration is larger than the Debye acceleration, i.e. $g>g_D$.
In other words, the limit of high temperatures compared to the
Debye temperature, is corresponding to the strong gravitational
fields. In this case we have $T\gg T_D$, thus for $x$ and $y$ in
the definition of the Debye function (\ref{Dx}), we have $x\ll1$
and consequently $y\ll1$. Therefore we can use the approximation
$e^y \approx 1+y$ in the integral of Eq. (\ref{Dx}) and as a
result, the one dimensional Debye function reduces to
\begin{equation}\label{DxVLT}
D(x) \approx \frac{1}{x}\int_0^x dy=1.
\end{equation}
Substituting this result ($D(x)=1$) in the modified Einstein
equations (\ref{modifiedEinstein}), leads to
\begin{equation}\label{standardEinstein}
R_{ab}=8\pi G (\mathcal{T}_{ab}-\frac{1}{2}\mathcal{T} g_{ab}).
\end{equation}
Therefore, in the temperatures extremely larger than the Debye
temperature (very strong gravitational fields), one obtains the
standard Einstein field equations as expected. }

\section{MOND theory from Debye entropic gravity}
Modified Newtonian dynamics (MOND) was proposed to explain the
flat rotational curves of spiral galaxies. A great variety of
observations indicate that the rotational velocity curves of all
spiral galaxies tend to some constant value \cite{Trimble}. Among
them are the Oort discrepancy in the disk of Milky Way
\cite{Bahcall}, the velocity dispersions of dwarf Spheroidal
galaxies \cite{Vogt} and the flat rotation curves of spiral
galaxies \cite{Rubin}. These observations are in contradiction
with the prediction of Newtonian theory because Newtonian theory
predicts that objects that are far from the galaxy center have
lower velocities.

The most widely adopted way to resolve these difficulties is the
dark matter hypothesis. It is assumed that all visible stars are
surrounded by massive nonluminous matters. Another approach is the
MOND theory which was suggested by M. Milgrom in 1983
\cite{Milgrom}. This theory appears to be highly successful for
explaining the observed anomalous rotational-velocity. {In fact,
the MOND theory is (empirical) modification of Newtonian dynamics
through modification in the kinematical acceleration term `$a$'
(which is normally taken as $a=v^{2}/r$ ) as effective kinematic
acceleration} $a_{\rm eff}=a \mu(\frac{a}{a_{0}})$,
\begin{equation}
\label{MOND} a \mu (\frac{a}{a_0})=\frac{G M}{R^2},
\end{equation}
{where $ \mu=1$ for usual-values of accelerations and $\mu=\frac{a}{%
a_{0}}$($\ll 1$) if the acceleration `$a$' is extremely low, lower
than a critical value $a_{0}=10^{-10}$ $m/s^{2}$. At large
distance, at the galaxy out skirt, the kinematical acceleration
`$a$' is extremely small, smaller than $10^{-10}$ $m/s^{2}$ ,
i.e., $a\ll a_{0}$, hence the function $\mu
(\frac{a}{a_{0}})=\frac{a}{a_{0}}$. Consequently, the velocity of
star on circular orbit from the galaxy-center is constant and does
not depend on the distance; the rotational-curve is flat, as it
observed.}

{Although MOND theory can explain the flat rotational curve,
however its theoretical origin remains un-known. Thus, it is well
motivated to establish a gravitational theory which can results
MOND theory naturally. In this section,  we are able to show that
the MOND theory can be extracted completely from the Debye
entropic gravity. This derivation further support the viability of
Debye entropic gravity formalism.}

Again, we consider a spherical holographic screen with radius $R$
as the boundary of the system. Combining Eqs. (3) and (4), and
using the equivalence between mass and energy as well as relation
$A=4\pi R^2$, we obtain
\begin{equation} \label{TDx}
\frac{2 \pi}{\hbar}T D(x)=\frac{G M}{R^2}.
\end{equation}
Using the Unruh temperature formula (\ref{TUnruh}), the above
equation may be written as
\begin{equation} \label{aDx}
a D(x)=\frac{G M}{R^2}.
\end{equation}
Also, if we use the Unruh temperature formula in the definition of
$x$, i.e. Eq. (\ref{x}), and define $a_0$ as
\begin{equation} \label{a0}
a_0 \equiv \frac{12 T_D}{\pi \hbar},
\end{equation}
then we obtain
\begin{equation} \label{x&a}
x=\frac{\pi^2 a_0}{6a}.
\end{equation}
Using the above result in Eq. (\ref{aDx}) gives
\begin{equation} \label{DebyeMOND}
a D(\frac{\pi^2 a_0}{6a})=\frac{G M}{R^2}.
\end{equation}
This is the MOND theory resulting from Debye entropic gravity. If
we compare this equation with well-known Eq. (\ref{MOND}), we see
that we can define $\mu$ function as
\begin{equation} \label{mu&D}
\mu(\frac{a}{a_0}) \equiv D(\frac{\pi^2 a_0}{6a}).
\end{equation}
In what follows we show that this function satisfies the
conditions similar to those of $\mu$ function in Eq. (\ref{MOND}).
Let us examine Eq. (\ref{DebyeMOND}) in two limits of
temperatures. First,
 we consider the limit corresponding to the temperatures large relative to the Debye temperature.
 In this case $x\ll 1$ ($a\gg a_0$) we have $D(x)=1$. Thus Eq.
 (\ref{DebyeMOND}) reduces to
\begin{equation}\label{DebyeMOND VLT}
a = \frac{GM}{R^2}.
\end{equation}
Therefore, for strong gravitational fields, Eq. (\ref{DebyeMOND})
turns into the standard Newtonian dynamics. As we discussed, for
$a\gg a_0$ we have also $\mu (\frac{a}{a_0})= 1$. We conclude that
in the limit of $a\gg a_0$ both $D(x)$ and $\mu(x)$ have the same
behavior and become equal to 1.

{The second limit corresponds to the temperatures extremely
smaller than the Debye temperature, $T\ll T_D$, that is to say in
the weak gravitational fields. In this limit, we have $x\gg1$
($a\ll a_0$), and the Debye function can be expanded as
\begin{equation} \label{DxST}
D(x) = \frac{1}{x} \int_0^\infty \frac{y}{e^y-1}
dy\approx\frac{\pi^2}{6x}.
\end{equation}
If we use the approximation (\ref{DxST}) in Eq. (\ref{DebyeMOND}),
we obtain
\begin{equation}\label{DebyeMOND ST}
a\left(\frac{a}{a_0}\right)=\frac{G M}{R^2}.
\end{equation}}
{Therefore, the Newtonian dynamics is modified for weak
gravitational fields, e.g. at large distance from the galaxy
center, namely at the galaxy out skirt. Thus the origin of the
MOND theory can be understood completely in the framework of Debye
entropic gravity. In this way we fill in the gap existing in the
literature understanding the theoretical origin of MOND theory.}
\section{Poisson equation from Debye entropic force}
Finally, we obtain the modified Poisson equation by taking into
account the Debye correction to the equipartition law of energy.
We choose a holographic screen $S$ corresponding to an
equipotential surface with fixed Newtonian potential $\phi_0$. We
assume that the entire mass distribution given by $\rho (\vec{x})$
is contained inside the volume enclosed by the screen and there
are some test
 particles outside this volume. To identify the temperature of the holographic screen, we take a
 test particle and move it close to the screen and measure its local acceleration. The local
  acceleration is related to the Newton potential as
\begin{equation} \label{a&phi}
\vec{a}=-\vec{\nabla}\phi.
\end{equation}
Substituting this relation into Unruh temperature formula, we get
\begin{equation} \label{T&phi}
T=\frac{\hbar |\vec{\nabla} \phi|}{2\pi}.
\end{equation}
Using the above equation in the definition of $x$, we  have
\begin{equation} \label{x&phi}
x\equiv\frac{T_D}{T}=\frac{2\pi T_D}{\hbar |\vec{\nabla} \phi|}.
\end{equation}
Inserting (\ref{T&phi}) in  Eq. (\ref{M&T}), after using Eq.
(\ref{N&A}) for the number of bits on the holographic screen, we
obtain
\begin{equation}\label{M&phi}
M=\frac{1}{4 \pi G}\oint_S D(x) \vec{\nabla} \phi . d\vec{A}.
\end{equation}
Using the divergence theorem we can rewrite Eq. (\ref{M&phi}) as
\begin{equation}\label{M&phi 2}
M=\frac{1}{4\pi G}\int_V \vec{\nabla} . [ D(x) \vec{\nabla} \phi
]dV.
\end{equation}
On the other hand, for the mass distribution $M$ inside the closed
surface $S$, we have the relation
\begin{equation}\label{M&rho}
M=\int_V \rho (\vec{x}) dV.
\end{equation}
Equating Eqs. (\ref{M&phi 2}) and (\ref{M&rho}), we get
\begin{equation}\label{Poisson}
\vec{\nabla} .[ D(x) \vec{\nabla} \phi ] = 4 \pi G \rho (\vec{x}).
\end{equation}
This is the modified Poisson equation which is valid in all range
of temperatures. For high temperatures. i.e. strong gravitational
field ($x\ll1$) and hence $D(x)=1$. In this case Eq.
(\ref{Poisson}) reduces to the standard Poisson equation,
\begin{equation}\label{PoissonVLT}
\nabla^2\phi = 4 \pi G \rho (\vec{x}).
\end{equation}
{Thus, considering the gravitational system as a thermodynamical
system and taking into account the Debye model for the modified
equipartition law of energy, we see that not only Einstein
equation and MOND theory but also the Poisson equation is modified
accordingly. Clearly the modification of Poisson equation leads to
modified Newton's law of gravitation.}
\section{Conclusions\label{Sum}}
In his work, Verlinde applied the equipartition law of energy as
$E=\frac{1}{2}NT$ on the holographic screen induced by the mass
distribution of the system, and obtained the Einstein equations,
Newton's law of gravitation and the Poisson equation. But we know
from statistical mechanics that the equipartition law of energy
does not hold at very low temperatures and it should be corrected.
In this paper, we considered the Debye correction to the
equipartition law of energy  as $E=\frac{1}{2}NTD(x)$, where
$D(x)$ is the Debye function. Following Verlinde's strategy on the
entropic origin of gravity, we obtained the modified form of the
Einstein equations, MOND theory  and the modified Poisson
equation. {Interestingly enough, we found that the origin of MOND
theory can be understood from the Debye entropic gravity scenario.
Since the MOND theory is an acceptable theory for explanation of
the galaxy flat rotation curves, thus the studies on its
theoretical origin is of great importance. This result is
impressive and show that the approach here is powerful enough for
deriving the modified gravitational field equations from Debye
model.} We also showed that in the temperatures extremely larger
than the Debye temperature (very strong gravitational fields), the
obtained modified equations turn into their respective well-known
standard equations. {The results obtained here further support the
viability of Verlinde's formalism.}
\acknowledgments{This work has been supported financially by
Center for Excellence in Astronomy and Astrophysics of IRAN
(CEAAI- RIAAM) under research project No. 1/2782-77.}

\end{document}